\documentclass[twocolumn,showpacs,preprintnumbers,amsmath,amssymb,fleqn,russian]{revtex4}
\usepackage{mathtext}
\usepackage{graphicx,epsfig}
\begin{document}
%\sloppy
\title{Scattering of TE- and TM-polarized wave packets on a dielectric layer: true and false group-delay times}
\author{N. L. Chuprikov\\ Tomsk State Pedagogical University,
\\634041, Tomsk, Russia}
\date{\today}

\begin{abstract}
We consider an oblique incidence on a uniform dielectric layer of the plane monochromatic TE- and TM-waves, as well as TE- and TM-polarized wave packets consisting of waves with
the same angle of incidence. For each polarization the stationary model is presented, which allows one to uniquely restore the dynamics of the transmitted and reflected
components of the incident wave packet at all stages of scattering. On this basis we introduce the concepts of ``true group-delay time"\/ and ``false group-delay time"\/ (the
corresponding quantities for transmission and reflection are the same). The former is defined as the difference of two instants in the evolution of the same wave packet, one of
which cannot be measured directly. Conversely, the latter is defined as the difference of two instants in the evolution of different wave packets, both can be measured directly.
The false group-delay time does not in itself have a physical sense, but plays a key role for an indirect measurement of the true group-delay times of both polarizations.
\end{abstract}
\maketitle

\newcommand{\Api}{A^{in}}
\newcommand{\Ami}{B^{in}}
\newcommand{\Apo}{A^{out}}
\newcommand{\Amo}{B^{out}}
\newcommand{\bpi}{a^{in}}
\newcommand{\bmi}{b^{in}}
\newcommand{\bpo}{a^{out}}
\newcommand{\bmo}{b^{out}}
\newcommand{\api}{a^{in}}
\newcommand{\ami}{b^{in}}
\newcommand{\apo}{a^{out}}
\newcommand{\amo}{b^{out}}
\newcommand {\uta} {\tau_{tr}}
\newcommand {\utb} {\tau_{ref}}

\newcommand{\FF}{\chi_{(1,j)}}
\newcommand {\aro}{(k)}
\newcommand {\da}{\partial}
\newcommand{\ppp}{\mbox{\hspace{5mm}}}
\newcommand{\ppa}{\mbox{\hspace{15mm}}}
\newcommand{\ppb}{\mbox{\hspace{20mm}}}
\newcommand{\ooo}{\mbox{\hspace{3mm}}}
\newcommand{\ooa}{\mbox{\hspace{1mm}}}

\section{Introduction} \label{a20}

\hspace*{\parindent} In classical electrodynamics, the concept of group velocity, as the velocity of the maximum of a wave packet, is successfully used to describe the
propagation of wave packets in various infinite media. However, all attempts to define on its basis the time it takes for the wave-packet maximum to pass through a layered
structure (this quantity is often referred to as ``phase time"\/, but we will call it ``group time"\/ because it is associated with the group velocity and not with the phase
velocity) proved to be unsatisfactory even in the case of the simplest structure -- a uniform dielectric layer, in which there is neither dissipation nor dispersion. Firstly,
there is no unambiguous definition of group time for this structure (see \cite{Gha,Ste,Lee}, as well as reviews \cite{Win} (see Exp. (34)) and \cite{Shw} (see Exp. (2.5)));
secondly, all the known definitions lead to the abnormally short or even negative phase time for transmission.

As is known, in connection with this, attempts were made to determine the speed of propagation of a wave packet through a layer not as the speed of the maximum of the wave
packet, but as the speed of its leading edge, discontinuity or precursor. However, no less difficult problems arose on this path, and there is reason to believe that the source
of these problems lies not in the very concept of group time, but in those versions of the stationary phase approximation on the basis of which it was defined. There are two
reasons why these versions should be considered unsatisfactory.

The first reason is related to the well-known fact that, in the course of scattering, the incident wave packet splits into two components: the transmitted and the reflected wave
packets. In this case, there is no complete cause-effect relationship between the incident wave packet and its transmitted component (see also \cite{But,But1}) because the
latter is only a part of the former. Thus, the maximum of the incident packet, its leading edge, discontinuity (if any), and precursor do not turn into the maximum, leading
front, discontinuity, and precursor of the transmitted packet, respectively. In the well-known definitions of the group time, this fact is simply ignored and this quantity is
derived from the comparison of the dynamics of the transmitted wave packet with that of the incident wave packet. The group time introduced in this way will be referred to as
``false group time"\/, since this conception a priori violates the causality principle. Thus, in the current model of scattering of electromagnetic pulses on a layered
structure, which does not imply an individual description of the transmitted and reflected components of the incident wave packet at the initial stage of scattering, one can
define only the false group time.

Another reason is manifested in the case of an oblique incidence of the wave packet on the layer, and we will discuss this in detail in the section \ref{Ch}. Now we want only to
point out the fact that at present there is no a generally accepted two-dimensional version of the stationary phase method and, as a consequence, there is no a generally
accepted definition of (false) group time for an oblique incidence (at the same time, as will be shown later, a reliable determination of this quantity is an important task,
since it plays a key role in the procedure of indirect measurement of true group time).

Our goal is to introduce group time based on a new two-dimensional version of the stationary phase method and the alternative model \cite{Chu} of scattering the TE-wave on a
dielectric layer. This model allows one to unambiguously reconstruct the dynamics of the transmitted and reflected wave packets at all stages of scattering. Since some
intermediate calculations, important for understanding the main idea of this approach, have been missed in the short article \cite{Chu}, they are presented in Section \ref{a11}
of the present paper. The standard scattering model (SSM) for the TE-wave (see p. 77-80 in \cite{Born}) is presented in Section \ref{a1}. In Section \ref{TM} we indicate the
changes that need to be made to extend this approach to the case of the TM-wave. In Section \ref{packet} we propose a new two-dimensional version of the stationary phase method
and gives definitions of group times for TE and TM polarized wave packets. An analysis of the known definitions of this quantity and the two-dimensional version \cite{Ste} of
the stationary phase method is presented in Section \ref{Ch}.

\section{Standard approach} \label{a1}

Following \cite{Born}, we consider two homogeneous nonmagnetic media ($\mu=1$) with dielectric constant values $\epsilon_0$ and $\epsilon$; dispersion and dissipation are
absent. The layer $a\leq z \leq b$ is filled with medium with an absolute refractive index $n$ ($n=\sqrt{\epsilon})$, and areas outside this layer are filled with vacuum or
medium with a refractive index $n_0$ ($n_0=\sqrt{\epsilon_0})$ close to one; $n>n_0$; $b-a=d$. The plane monochromatic TE-wave is incident from the left onto the layer; the unit
vector defining its direction of incidence lies in the $YZ$ plane and forms an angle $\theta$ with the $OZ$ axis.

In this case, the component $E_x$ of the electric field and two components, $H_y$ and $H_z$, of the magnetic field are non-zero. Each of these three components satisfies the
wave equation whose solution is sought in the form:
\[E_x=U(z)e^{i\chi},\ooo H_y=V(z)e^{i\chi},\ooo H_z=W(z)e^{i\chi};\]
\[\chi=k (n_{y} y-c t),\ooo n_{y}=n_0\sin\theta;\] $c$ is the speed of light in vacuum. The searched-for real components of the fields are $\Re(Ue^{i\chi})$, $\Re(Ve^{i\chi})$
and $\Re(We^{i\chi})$. For the energy density $w$ and the energy flux density $\textbf{S}$ of the TE-wave in a medium with a dielectric constant $\epsilon$, we have
\begin{eqnarray*}
w=w^{(0)}+w^{(t)},\ooo \textbf{S}=\textbf{S}^{(0)}+\textbf{S}^{(t)},
\end{eqnarray*}
\begin{eqnarray*}
w^{(0)}(z)= \frac{1}{16\pi}\left(\epsilon |U|^2+|V|^2+|W|^2\right);\\ w^{(t)}(y,z)=\frac{1}{16\pi}\Re\left[\left(\epsilon
U^2+V^2+W^2\right)e^{2i\chi}\right];
\end{eqnarray*}
$S_x^{(0)}=S_x^{(t)}=0$;
\begin{eqnarray*}
S_y^{(0)}=-\frac{c}{8\pi}\Re\left(U^*W\right);\ooa S_y^{(t)}=-\frac{c}{8\pi}\Re\left(UWe^{2i\chi}\right)
\end{eqnarray*}
\begin{eqnarray*}
S_z^{(0)}= \frac{c}{8\pi}\Re\left(U^*V\right);\ooo S_z^{(t)}=\frac{c}{8\pi}\Re\left(UVe^{2i\chi}\right).
\end{eqnarray*}
Quantities with the index $(0)$ depend only on $z$.

The functions $V(z)$ and $W(z)$ are associated with $U(z)$ (see \cite{Born})
\begin{eqnarray}\label{5}
V(z)=-iU^\prime(z)/k,\ooo W(z)=-U(z)n_{y};
\end{eqnarray}
the prime denotes the derivative with respect to $z$. In this case
\begin{eqnarray} \label{10c}
w^{(0)}(z)=\frac{1}{16\pi}\left[\left(n^2+n_{y}^2\right)|U(z)|^2+\frac{|U'(z)|^2}{k^2}\right]\\
S_y^{(0)}=\frac{c n_{y}}{8\pi}|U|^2,\ooo S_z^{(0)}=\frac{c}{8\pi k}\Im(U^*U').\nonumber
\end{eqnarray}
That is, it suffices to find the function $U(z)$.

Outside and within the interval $[a,b]$, the original wave equation for the electric field is reduced, respectively, to the equations (see \cite{Born})
\begin{eqnarray}\label{6}
U^{\prime\prime}+k^2 n_{z}^2U=0,\ooo U^{\prime\prime}+k^2(n^2-n_{y}^2)^2 U=0;
\end{eqnarray}
where $n_{z}=n_0\cos\theta$.

At the boundaries of this interval, the function $U(z)$ and its first derivative $U^\prime(z)$ must be continuous. This follows from the continuity condition for the tangential
projections $E_x$ and $H_y$ and the normal projection $H_z$, as well as from the relations (\ref{5}). Note that for the TE-wave, in addition to the electromagnetic energy
conservation law that follows from the continuity equation
\begin{eqnarray*}
\frac{\partial w}{\partial t} +\nabla \textbf{S}=0,
\end{eqnarray*}
we also have a conservation law
\begin{eqnarray*}
S_z^{(0)}(z)= \frac{c}{8\pi k}\Im\left(U^*U^\prime\right)=const.
\end{eqnarray*}

The solution of Eqs. (\ref{6}) can be written in the form (here we introduce notations that will be used in section \ref{a11}):
\begin{eqnarray} \label{7}
U(z)=\left\{
\begin{array}{rl}
U_{tot}^{inc}(z)+U_{tot}^{ref}(z),\ooo z<a\\
C^{(1)}_{tot} F_1(z)+C^{(2)}_{tot} F_2(z),\ooo a<z<b\\ U_{tot}^{tr}(z),\ooo z>b
\end{array} \right.
\end{eqnarray}
\begin{eqnarray*}
U_{tot}^{inc}(z)=e^{ikn_{z}z},\ooo U_{tot}^{ref}(z)=b_{out}(k) e^{ikn_{z} (2a-z)},
\end{eqnarray*}
\begin{eqnarray*}
U_{tot}^{tr}(z)= a_{out}(k) e^{ikn_{z}(z-d)}.
\end{eqnarray*}
Inside the layer
\begin{eqnarray*}
F_1(z)=\sin[\mathcal{N}k(z-z_c)],\ooa F_2(z)= \cos[\mathcal{N}k(z-z_c)]
\end{eqnarray*}
$\mathcal{N}=\sqrt{n^2-n_{y}^2}$, $z_c=(b+a)/2$.

By matching the solutions on the boundaries of the layer, we obtain the equations for $a_{out}$, $b_{out}$, $C^{(1)}_{tot}$ è $C^{(2)}_{tot}$:
\begin{eqnarray*}
\mathcal{N} C^{(1)}_{tot} e^{-ikn_z a}=P+b_{out}P^*=-a_{out} P^*;\\
\mathcal{N} C^{(2)}_{tot}e^{-ikn_z a}=Q+b_{out}Q^*=a_{out}Q^*;
\end{eqnarray*}
where, with $\phi=\mathcal{N} kd/2$,
\begin{eqnarray*}
Q=\mathcal{N}\cos(\phi)+in_{z}\sin(\phi),\\ P=-\mathcal{N}\sin(\phi)+in_{z}\cos(\phi).
\end{eqnarray*}
From which it follows that
\begin{eqnarray} \label{10b}
a_{out}=\frac{1}{2}\left[\frac{Q}{Q^*}-\frac{P}{P^*}\right],\ooa b_{out}=-\frac{1}{2}\left[\frac{Q}{Q^*}+\frac{P}{P^*}\right] \\
C^{(1)}_{tot}e^{-ikn_z a}=-\frac{P^*}{\mathcal{N}} a_{out} \equiv \frac{in_z}{Q^*},\nonumber\\ C^{(2)}_{tot}e^{-ikn_z a}=\frac{Q^*}{\mathcal{N}} a_{out}\equiv-\frac{in_z}{P^*}.
\end{eqnarray}

The amplitudes $a_{out}$ and $b_{out}$ can be written as
\begin{eqnarray}  \label{10c}
a_{out}=\sqrt{T} e^{iJ},\ooo b_{out}=-i\sqrt{R} e^{iJ},
\end{eqnarray}
where $T$ and $R$ are the transmission and reflection coefficients, respectively; $T+R=1$:
\begin{eqnarray} \label{11}
T=\frac{1}{1+\eta_{(-)}^2 \sin^2(\mathcal{N} kd)}\equiv \nonumber\\  \frac{1}{\left[\eta_{(+)}-\eta_{(-)}
\cos(\mathcal{N} kd)\right]\left[\eta_{(+)}+\eta_{(-)} \cos(\mathcal{N} k d)\right]} \nonumber\\
J=\arctan\left[\eta_{(+)}\tan(\mathcal{N} k d)\right],\ooa \eta_{(\pm)}=\frac{\mathcal{N}^2\pm n_z^2}{2\mathcal{N}n_z}
\end{eqnarray}
It is easy to check that $S_z^{(0)}=c n_{z}T/(8\pi)$.

\section{TE-wave: transmitted and reflected components} \label{a11}

We have to stress that on the basis of the SMR, allowing one to investigate the throughput and reflectivity of the structure under study,  it is not impossible, in principle, to
determine the time of energy transfer of that component of the TE-wave that passes through the layer, as well as that its component that is reflected from this layer. This is
due to the fact that the expression in (\ref{7}), for the TE-wave inside the layer, does not imply an individual description of its components.

From the point of view of the principle of causality, there must be an incident wave $U_{tr}^{inc}(z)$ in the region $z<a$ that has a cause-effect relationship with the
transmitted wave $U_{tot}^{tr}(z)$, as well as the incident wave $U_{ref}^{inc}(z)$, which has a cause-effect relationship with the reflected wave $U_{tot}^{ref}(g)$. In other
words, there should exist a sole pair of functions $U_{tr}(z)$ and $ U_{ref}(z)$, which are uniquely determined by the function $U(z)$: the first one describes that component of
the scattering TE-wave, which ultimately passes through the layer, and the second is that component that is reflected off it.

All requirements for the functions $ U_ {tr} (z) $ and $ U_ {ref} (z) $ arising from the superposition and causality principles can be formulated as follows:

(a) the scattering TE-wave should be a superposition of its transmitting and reflecting components: $U(z)=U_{tr}(z)+U_{ref}(z)$ for any $z$ on the $OZ$ axis;

(b) since each of these TE-wave components represents only one scattering channel, the functions $U_{tr}(z)$ and $U_{ref}(z)$ must have one incident wave and one outgoing wave;
in this case
\begin{eqnarray*}
U_{tr}(z)=\left\{
\begin{array}{rl}
U_{tr}^{inc}(z);\ooo z<a\\
U_{tot}^{tr}(z);\ooo z>b
\end{array} \right.
\end{eqnarray*}
\begin{eqnarray*}
U_{ref}(z)=\left\{
\begin{array}{rl}
U_{ref}^{inc}(z)+U_{tot}^{ref}(z);\ooa z<a\\
0;\ooa z>b
\end{array} \right.
\end{eqnarray*}

(c) there is a plane $z_0$ inside the layer, where the continuation into the layer region of the incident wave $U_{tr}^{inc}(z)$ is ``stitched" with that of the outgoing wave
$U_{tot}^{tr}(z)$; on this plane, the function $U_{tr}(z)$ (and $U_{ref}(z)$) must be continuous, together with the vectors $\textbf{S}_{ref}^{(0)}(z)$ and
$\textbf{S}_{tr}^{(0)}(z)$.

The choice of this continuity condition instead of the "ordinary" one is related to the fact that among the solutions of Eqs. \ref{6} for a partially transparent layer there are
no functions that would be continuous on the plane $ z_0 $, along with their first derivatives, and also possess only one incident wave and only one outgoing wave. That is why
the first derivative with respect to $z$ of each of the functions $U_{tr}(z)$ and $U_{ref}(z)$ must satisfy on this plane a weaker continuity condition -- the continuity of the
vectors $\textbf{S}_{ref}^{(0)}(z)$ and $\textbf{S}_{tr}^{(0)}(z)$. Considering the fact that the corresponding energy densities are continuous on this plane, these continuity
conditions provide a cause-effect relationship between the incoming and outgoing waves for each of the functions $U_{tr}(z)$ and $U_{ref}(z)$.

Before we start searching for these functions, we should note the following. From the condition (c) it follows that $U_{tr}(z)\equiv U_{tot}(z)$ for $z\geq z_0$. This means that
in this region $U_{ref}(z)\equiv 0$, and the $z$-th projection $\textbf{S}_{ref}^{(0)}(z)$ is identically zero on the $OZ$-axis. Obviously, $z_0=z_c$ for a homogeneous layer,
since the solution inside the layer (see (\ref{7})) is expressed in this case through even and odd (relative to the $z=z_c$ plane) functions. The expression for the function
$U_{ref}(z)$ inside the layer should contain only the odd function $F_1(z)$.

Note also that for sufficiently large values of $k$, the function $U_{ref}(z)$ can have many zeros inside the interval $[0,z_c)$. At first glance, in this case we would have to
take as $z_0(k)$ that zero which is closest to the left boundary of the layer. However, the TE-wave under study is of interest not on its own, but as the main harmonic of the
(strictly speaking, infinitely)  narrow  in $k$-space wave packet scattering on the layer. Therefore, as $z_0(k)$, we must take that zero of the function$U_{ref}(z)$ which is
independent on $k$. Thus,
\begin{eqnarray} \label{70}
U_{tr}=\left\{
\begin{array}{rl}
A_{tr} e^{ikn_{z}z},\ooa z<a\\
C^{(1)}_{tr} F_1(z)+C^{(2)}_{tot} F_2(z),\ooa a<z<z_c,\\
U(z),\ooa z>z_c
\end{array} \right.\nonumber\\
U_{ref}=\left\{
\begin{array}{rl}
A_{ref} e^{ikn_{z}z}+b_{out} e^{ikn_{z} (2a-z)},\ooa z<a\\
C^{(1)}_{ref} F_1(z),\ooa a<z<z_c,\\
0,\ooa z>z_c
\end{array} \right.
\end{eqnarray}

By "stitching" on the plane $z=a$ of Exps. (\ref{70}) obtained for the function $U_{ref}(z)$ in the regions $z<a$ and $a<z<z_c$, we get
\begin{eqnarray} \label{71}
A_{ref}=b_{out}^*(a_{out}+b_{out})\equiv -b_{out}(a^*_{out}-b^*_{out})\nonumber\\
C^{(1)}_{ref}=-2 b_{out}\frac{Q^*}{Q}\ooa C^{(1)}_{tot}
\end{eqnarray}
Since $A_{tr}+A_{ref}=1$ and $C^{(1)}_{tr}+C^{(1)}_{ref}=C^{(1)}_{tot}$, for the transmitted component we get
\begin{eqnarray} \label{72}
A_{tr}=a_{out}(a_{out}^*-b_{out}^*)\equiv a^*_{out}(a_{out}+b_{out}),\nonumber\\
C^{(1)}_{tr}=-\frac{PQ^*}{P^*Q}\ooa C^{(1)}_{tot}.
\end{eqnarray}
$A_{tr}$ and $A_{ref}$ can also be written in the form
\begin{eqnarray} \label{72a}
A_{tr}=T+i\rho\sqrt{T R}\equiv\sqrt{T} \exp(i\lambda),\nonumber\\
A_{ref}=R-i\rho\sqrt{R T}\equiv\sqrt{R} \exp\left[i\left(\lambda-\rho \frac{\pi}{2}\right)\right],\\
\lambda=\arctan\left(\rho\sqrt{\frac{R}{T}}\right),\ooo \rho=sign[\sin(k\mathcal{N}d)]. \nonumber
\end{eqnarray}

Let us now dwell on the basic properties of the functions $U_{tr}(z)$ and $U_{ref}(z)$, which follow from the expressions (\ref{70})-(\ref{72a}). First, the energy density
\begin{eqnarray} \label{74}
w^{(0)}_{tr}=\frac{1}{16\pi}\left[\left(n^2+n_{y}^2\right)|U_{tr}(z)|^2+\frac{|U'_{tr}(z)|^2}{k^2}\right]
\end{eqnarray}
and energy flux density
\begin{eqnarray} \label{75}
\mathbf{S}^{(0)}_{tr}=\left(0, \frac{c n_{y}}{8\pi}|U_{tr}(z)|^2, S_z^{(0)}\right)
\end{eqnarray}
are continuous on the $z=z_c$ plane. This follows from the equality $|C^{(1)}_{tr}|=|C^{(1)}_{tot}|$. That is, despite the fact that the derivative of $U'_{tr}(z)$ is
discontinuous on the plane $z=z_c$, its absolute value $|U^\prime_{tr}(z)|$ on this plane is continuous, insofar as
\begin{eqnarray} \label{149}
|U_{tr}(z_c-z)|=|U_{tr}(z-z_c)|,\nonumber\\ |U'_{tr}(z_c-z)|=|U'_{tr}(z-z_c)|.
\end{eqnarray}

As for the tangential components $E_x^{tr}$ and $E_x^{ref}$ of the electric field and the normal components $H_z^{tr} $ and $H_z^{ref}$ of the magnetic field, they are
continuous on the plane $z=z_c$. But the tangential components $H_y^{tr}$ and $H_y^{ref}$ for the transmitted and reflected components of the original wave are discontinuous
here. The discontinuity of $H_y^{tr}(z)$ results from the electric current, in the plane $z=z_c$, associated with the $U_{tr}$ component; and the discontinuity of $H_y^{ref}(z)$
results from the electric current on this plane associated with $U_{ref}$. But all these currents exist indirectly; just like the components themselves, which cannot be
separated from each other. And, since the total current of both components on the plane $z=z_c$ is zero, the total field $H_y^{tr} H_y^{ref}=H_y$ has no real discontinuities on
this plane.

Secondly, as it follows from Exps. (\ref{72a}), not only $A_{tr}+ A_{ref}=1$, but also $|A_{tr}|^2+|A_{ref}|^2=1$. Thus, the superposition of the incident waves
$U_{tr}^{inc}(z)$ and $U_{ref}^{inc}(z)$ as well as the sum of the corresponding energy densities (both are conserved in the course of scattering) give the incident TE-wave
$U_{tot}^{inc}(z)$ and the energy density $\left(w^{inc}_{tot}\right)^{(0)}$, respectively:
\begin{eqnarray} \label{73}
U_{tr}^{inc}(z)+U_{ref}^{inc}(z)=U_{tot}^{inc}(z),\nonumber\\
\left(w^{inc}_{tr}\right)^{(0)}+\left(w^{inc}_{ref}\right)^{(0)}=\left(w^{inc}_{tot}\right)^{(0)}=\frac{n^2+n_0^2}{16\pi}.
\end{eqnarray}

Thus, the wave $U_{ref}^{inc}(z)=A_{ref} e^{ikn_{z}z}$ really describes that component of the original wave $e^{ikn_{z}z}$ which is ultimately reflected off the layer, while
$U_{tr}^{inc}(z)=A_{tr}e^{ikn_{z}z}$ describes the component which passes ultimately through it. Similar relations also arise for the wave packet $\mathcal{U}_{tot}(z,t)$,
$\mathcal{U}_{tr}(z,t)$ and $\mathcal{U}_{ref}(z,t)$ constructed from the corresponding stationary waves.

\section{Scattering of the TM-wave} \label{TM}

If one uses the known substitution rules, the results obtained for the TE-wave are easily transferred onto the case of scattering of the TM-wave. Indeed, while in the case of
the TE-wave, the nonzero components of the field vectors are $E_x$, $H_y$ and $H_z$, in the case of a TM-wave, these are $H_x$, $E_y$ and $E_z$. While in the first case, the
tangential component of the electric field and the normal component of the magnetic field, continuous at the layer boundaries regardless of the presence of electric charges and
currents at these boundaries, are associated with the function $U(z)$, in the second case they are associated with the derivative $U^\prime(z)$ (see p. 68 in \cite{Born}). The
remaining components are also continuous at these boundaries, since there are no electric charges and currents at these boundaries. As a result, the function $U(z)$ and the
transmission coefficient are the same for both waves.

But restoring the whole dynamics of the transmitted and reflected components of the original wave depends on its polarization. As we have shown, the plane $z=z_c$ plays a
special role in this case, and continuity on this plane can be guaranteed only for those components of field vectors that must be continuous at the boundaries of two media,
regardless of the presence or absence on them of electric charges and currents. In the case of the TE-wave, this condition is fulfilled when the function $U_{ref}$ is continuous
and equal to zero on the plane $z=z_c$ -- the choice of this function guarantees the continuity of the components $E_x$ and $H_z$ on the plane $z=z_c$ for the reflected and
transmitted component of the original wave. In the case of the TM-wave,  this condition is fulfilled when the first derivative of the function $U_{ref}^{TM}$  is continuous and
zero on the $z=z_c$ plane. The needed function corresponds to the root
\begin{eqnarray} \label{72b}
\lambda=-\arctan\left(\rho\sqrt{R/T}\right):
\end{eqnarray}
\begin{eqnarray*}
U_{tr}^{TM}=\left\{
\begin{array}{rl}
A_{tr}^* e^{ikn_{z}z},\ooa z<a\\
C^{(1)}_{tot} F_1(z)+C^{(2)}_{tr} F_2(z),\ooa a<z<z_c\\
U(z),\ooa z>z_c
\end{array} \right.\nonumber\\
U_{ref}^{TM}=\left\{
\begin{array}{rl}
A_{ref}^* e^{ikn_{z}z}+b_{out} e^{ikn_{z} (2a-z)},\ooa z<a\\
C^{(2)}_{ref} F_2(z),\ooa a<z<z_c\\
0,\ooa z>z_c
\end{array} \right.\\
C^{(2)}_{tr}=\frac{P^*Q}{PQ^*}\ooa C^{(2)}_{tot},\ooo C^{(2)}_{ref}=-2 b_{out}\frac{P^*}{P}\ooa C^{(2)}_{tot}.
\end{eqnarray*}

It is easy to show that, for the found TE and TE-waves, not only the properties (\ref{149}) but also the properties (\ref{73}) hold. Since they are also applicable to the
corresponding wave packets, it is now possible to determine the group transmission and reflection times, as well as the corresponding delays, for both TE- and TM-polarized wave
packets.

\section{True and false group-delay times} \label{packet}

We have to begin with the name of this concept. First of all, we note that the timekeeping procedure proposed below gives the same result for the transmitted and reflected
components, therefore further we focus our attention on the transmitted component, and do not specify the channel of scattering in the name of group time. Another refinement
that is required here is due to the fact that in the well-known work \cite{Ste}, apart from the name ``phase time"\/ this concept is named the 11group-delay time". However, the
word ``delay"\/ in the second name, like the word ``phase"\/ in the first one, is misleading since, in the paper \cite{Ste}, it is defined the extrapolated (asymptotic) time of
dwelling inside the layer of the maximum of the transmitted component (see section \ref{Ch}), rather than the difference of this time and time of a free passage of this maximum
through this region. In the first case, it is appropriate to speak of ``group time"\/ and not of ``phase time"\/ or of ``group-delay time"; and only in the second case it is
appropriate to speak of ``group-delay time".

Consider the scattering of a wave packet built of the TE-waves with the same angle of incidence $\theta$,
\begin{eqnarray*}
\mathcal{U}_{tot}(y,z,t)=\int_0^\infty {\mathcal{A}}(k) U(z;k) e^{ik(n_0 y\sin\theta-c t)}dk,
\end{eqnarray*}
$\mathcal{A}(k)$ is a smooth real function decreasing, together with its derivatives, faster than any power of $1/k$, when $k\to \infty$, and faster than any power of $k$, when
$k\to 0$. Then the component of the wave packet $\mathcal{U}_{tot}$ passing through the layer is
\begin{eqnarray*}
\mathcal{U}_{tr}(y,z,t)=\int_0^\infty {\mathcal{A}}(k) U_{tr}(z;k) e^{ik(n_0 y\sin\theta-c t)}dk.
\end{eqnarray*}
In order to use the stationary phase approximation, we assume that $\mathcal{A}(k)$ gives a wave packet with a sufficiently small $\Delta k$ width in the $k$-space; the wave
number of the main harmonic of such a packet will also be denoted by $k$. Its width $\Delta l$ in the coordinate space ($\Delta l\sim 1/\Delta k$) satisfies the condition
$d\ll\Delta l\cos\theta\ll a$.

In accordance with our approach and the known stationary phase approximation in the problem of the scattering of the wave packet $\mathcal{U}_{tot}$ on a dielectric layer, the
group time describing its transmitted component should be determined from the analysis of the dynamics of the maximum of the wave packet $\mathcal{U}_{tr}$ at the initial and
final scattering stages. But the intrigue lies in the fact that the definition of this quantity also implies the knowledge of the dynamics, at the initial stage of scattering,
of the very wave packet $\mathcal{U}_{tot}$.

Since $\mathcal{A}(k)$ is a real function, the phases of the main harmonic of the $\mathcal{U}_{tr}$ component before and after scattering are described, respectively, by the
expressions (see also (\ref{70}), (\ref{72a}) and (\ref{7}))
\begin{eqnarray} \label{76}
\Phi^{inc}_{tr}=\lambda+kn_0[z\cos\theta+y\sin\theta]-ckt,
\end{eqnarray}
\begin{eqnarray} \label{77}
\Phi_{tr}^{out}=J+kn_0[(z-d)\cos\theta+y\sin\theta]-ckt
\end{eqnarray}
The main harmonic phase of the original $\mathcal{U}_{tot}$ packet at the initial scattering stage is
\begin{eqnarray} \label{76a}
\Phi^{inc}_{tot}=kn_0[z\cos\theta+y\sin\theta]-ckt.
\end{eqnarray}

Thus, in the framework of the stationary phase approximation, long before the scattering event, the dynamics of the maximum of the wave packet $\mathcal{U}_{tot}$  are described
by the equation
\begin{eqnarray} \label{78a}
\frac{d\Phi_{tot}^{inc}}{d k}=n_0[z\cos\theta+y\sin\theta]-ct=0,
\end{eqnarray}
while its component $\mathcal{U}_{tr}$ is described by the equation
\begin{eqnarray} \label{78}
\frac{d\Phi_{tr}^{inc}}{d k}=\lambda_k+n_0[z\cos\theta+y\sin\theta]-ct=0,
\end{eqnarray}
where $\lambda_k$ is the derivative of the function $\lambda(k)$. For this component, at the stage when scattering of the investigated part of the wave packet has been
completed, we have
\begin{eqnarray} \label{79}
J_k+n_0[(z-d)\cos\theta+y\sin\theta]-ct=0,
\end{eqnarray}
where $J_k$ is the derivative of the phase $J(k)$. Taking into account (\ref{11}), (\ref{72}) and (\ref{72a}), it can be shown that
\begin{eqnarray} \label{44}
J_k=\eta_{(+)} T \mathcal{N} d,\ppp \lambda_k=\eta_{(-)} T \mathcal{N} d \cos(\mathcal{N} kd).
\end{eqnarray}

According to the equations (\ref{78a}) and (\ref{78}), the maximum of the wave packet $\mathcal{U}_{tot}$ passes the origin at time $t^{st}_{tot}=0$ while the maximum of its
component $\mathcal{U}_{tr}$ is at the origin at the moment $t^{st}_{tr}=\lambda_k/c$. This confirms the fact that the peak of the incident wave packet $\mathcal{U}_{tot}$ does
not turn into the peak of the transmitted packet (see \cite{But,But1}). Nevertheless, the time point $t^{st}_{tot}=0 $ plays a key role in the timekeeping procedure of the
dynamics of the $\mathcal{U}_{tr}$ component -- in fact, this is the time of beginning the experiment which is fixed by the experimenter. As for the moment of time
$t^{st}_{tr}=\lambda_k/c$, which is determined from the scattering data, its direct measurement is impossible in principle (see the next section).

Note that far from the layer, the direction of motion of the maxima of the wave packets $\mathcal{U}_{tot}$ and $\mathcal{U}_{tr}$ is given by the angle $\theta$. Therefore, for
those infinitely small pieces of the maxima of both packets that pass through the origin, the variables $y$ and $z$ in Eqs. (\ref{78a}) and (\ref{78}) are related as follows
$y=z\tan\theta$. For these pieces
\begin{eqnarray*}
n_0z/\cos\theta-ct=0,\ooo \lambda_k+n_0z/\cos\theta-ct=0.
\end{eqnarray*}
From which it follows that if these pieces of the packets moved free in the whole space, they would arrive at the point, on the other side of the layer, with the coordinates
$y_{fin}=(a+d+L)\tan\theta$ and $z_{fin}=a+d+L$, at the instant of time
\begin{eqnarray*}
(t_{tot}^{fin})_{free}=\frac{n_0(a+d+L)}{c\cos\theta};\\
(t_{tr}^{fin})_{free}=\frac{\lambda_k}{c}+\frac{n_0(a+d+L)}{c\cos\theta},
\end{eqnarray*}
respectively; here $L$ is the distance from this point to the layer, and $L\gg\Delta l$. Therefore, at this spatial point, the equation (\ref{79}) holds, from which it follows
that the maximum of the transmitted wave packet passes through this point at the time
\begin{eqnarray*}
t_{fin}=\frac{1}{c}\left(J_k+n_0\frac{a+L}{\cos\theta}+n_0 d\tan\theta\sin\theta\right).
\end{eqnarray*}

The difference $t_{fin}-(t_{tr}^{fin})_{free}$ is the searched for (asymptotic) group-delay time $\Delta\tau_{TE}^{gr}$, which describes the passage of the component
$\mathcal{U}_{tr}$ through the dielectric layer:
\begin{eqnarray} \label{80}
\Delta\tau_{TE}^{gr}=\frac{1}{c}\left(J_k-\lambda_k-n_0d\cos\theta\right).
\end{eqnarray}
Adding the time $\tau_{free}=n_0d/(c\cos\theta)$ of free passage through the layer, we obtain the (asymptotic) group time (for transmission) $\tau_{TE}^{gr}$:
\begin{eqnarray} \label{81}
\tau_{TE}^{gr}=\frac{1}{c}\left(J_k-\lambda_k+n_0d\tan\theta\sin\theta\right).
\end{eqnarray}

For the time $\tau_{TE}^{gr}$, the maximum of the $\mathcal{U}_{tr}$ component shifts along the $y$-direction with the speed $c\sin\theta/n_0$ by $(\Delta
y)_{TE}^{gr}=\tau_{TE}^{gr}\cdot c\sin\theta/ n_0$. It is easy to show that
\begin{eqnarray} \label{83}
\frac{\tau_{TE}^{gr}}{\tau_{free}}=\frac{(\Delta y)_{TE}^{gr}}{(\Delta y)_{free}};
\end{eqnarray}
here $(\Delta y)_{free}=d\tan\theta$ is the lateral shift with the free passage of the layer.

Considering in the same manner the reflected component
\begin{eqnarray*}
\mathcal{U}_{ref}(y,z,t)=\int_0^\infty {\mathcal{A}}(k) U_{ref}(z;k) e^{ik(n_0 y\sin\theta-c t)}dk,
\end{eqnarray*}
it is easy to show that the (asymptotic) group time for reflection coincides with $\tau_{TE}^{gr}$.

As will be shown later, in this timekeeping procedure, the key role is played by the quantity $\Delta\tau^{gr}_{tot}=t_{fin}-(t_ {tot}^{fin})_{free}$, which we will call "false
group-delay time":
\begin{eqnarray} \label{80a}
\Delta\tau^{gr}_{tot}=\frac{1}{c}\left(J_k-n_0d\cos\theta\right).
\end{eqnarray}
This name reflects the fact that the initial instant of time that is included in the definition of this quantity does not describe the transmitted wave packet at the initial
instant of time. That is, $\Delta\tau^{gr}_{tot}$ cannot be interpreted as true group-delay time for the transmitted component.

However, both the time moment $t_{fin}$, and the time moment $(t_{tot}^{fin})_{free}$, as opposed to the moment $(t_{tr}^{fin})_{free}$ used in the definition (\ref{80}) can be
{\it directly} measured. Therefore $\Delta\tau^{gr}_{tot}$ plays a key role in the procedure of indirect measurement of the true group delay times for TE- and TM-polarized wave
packets. In addition, on the basis of measurements of this quantity, one can experimentally verify the timekeeping procedure itself.

\section{How to measure true group-delay time}

So, since the moment of time $t_{tr}^{st}=\lambda_k/c$, which is included in the definition of the true group delay time $\Delta\tau_{TE}^{gr}$, was found from the scattering
data, its direct measurement is impossible. This definition must be complemented by the formalism necessary for an indirect measurement of $\Delta\tau_{TE}^{gr}$. This task is
similar to solving the inverse scattering problem. The only difference is that in this case it is necessary to reconstruct, by the scattering data, not the properties of the
scatterer (this is, just, what is known), but the properties of the transmitted and reflected components of the incident packet, at the initial instant of time. In the
stationary phase approximation, this means that by the known transmission coefficients of $T$ and reflection of $R$, it is necessary to find the phases of the transmitted and
reflected components of the main harmonic $e^{ik(n_0 y\sin\theta + n_0 z\cos\theta- c t)}$ of the incident wave packet.

Let us consider the main harmonic of the incident wave packet $\mathcal{U}_{tot}$ at the initial moment of time at the origin (providing that the amplitude of the main harmonic
is equal to unity). Since its $T$-th part passes through the layer, and the $R$-th part is reflected, the passing and reflecting components of the main harmonic are equal in
this case, respectively, $\sqrt{T}e^{i\alpha}$ and $\sqrt{R}e^{i\beta}$. Here $\alpha$ and $\beta$ are unknown phases that obey the equation
\begin{eqnarray*}
\sqrt{T}e^{i\alpha}+\sqrt{R} e^{i\beta}=1.
\end{eqnarray*}

This equation has two roots:
\begin{eqnarray} \label{86}
\beta=\alpha-\frac{\pi}{2},\ooo \alpha=\arctan\left(\pm \rho\sqrt{\frac{R}{T}}\right),
\end{eqnarray}
The appearance of two roots is not accidental. In Sections \ref{a11} and \ref{TM} we showed that the first root describes the transmitted and reflected components of the TE-wave
(see (\ref{72a})), and the second root describes the TM-wave (see (\ref{72b})). (As is seen, only the scattering data are needed to search for the roots themselves. A detailed
analysis of the solutions inside the layer in Sections \ref{a11} and \ref{TM} was needed only to find out which root is associated with the TE-wave, and which root is associated
with the TM-wave.)

So, the root (\ref{86}) with a negative sign describes the TM-wave. That is, the true group-delay time $\Delta\tau_{TE}^{gr}$ for a TE-polarized wave packet is closely related
to the true group-delay time $\Delta\tau_{TM}^{gr}$ for a TM-polarized packet
\begin{eqnarray} \label{80b}
\Delta\tau_{TM}^{gr}=\frac{1}{c}\left(J_k+\lambda_k-n_0d\cos\theta\right),
\end{eqnarray}
which can also be measured only indirectly. We show that the procedure of indirect measurement of both quantities by the known false group-delay time (which is the same for both
polarizations) is unambiguous.

Let $L_{TE}=J_k-\lambda_k$ and $L_{TM}=J_k+\lambda_k$. Given (\ref{11}) and (\ref{44}), it is easy to show that
\begin{eqnarray} \label{88a}
L_{TE}=\frac{\mathcal{N} d}{\eta_{(+)}+\eta_{(-)} \cos(\mathcal{N} kd)},\nonumber\\
L_{TM}=\frac{\mathcal{N} d}{\eta_{(+)}-\eta_{(-)} \cos(\mathcal{N} kd)}.
\end{eqnarray}
The following equalities hold
\begin{eqnarray*}
L_{TE}+L_{TM}=2 J_k;\ooo L_{TE}\cdot L_{TM} =\min(J_k)\cdot J_k;
\end{eqnarray*}
$\min(J_k)=\mathcal{N}d/\eta_{(+)}$ (see (\ref{44}) and (\ref{11})).

The true relative delays $\mathcal{E}_{TE}=\Delta\tau_{TE}^{gr}/\tau_{free}$ and $\mathcal{E}_{TM}=\Delta\tau_{TM}^{gr}/\tau_{free}$ are associated with the false relative delay
$\mathcal{E}_{tot}=\Delta\tau^{gr}_{tot}/\tau_{free}$ by the relations
\begin{eqnarray} \label{88}
\mathcal{E}_{TE}+\mathcal{E}_{TM}=2 \mathcal{E}_{tot},\nonumber\\
\left(\mathcal{E}_{TE}+\cos^2(\theta)\right)\left(\mathcal{E}_{TM}+\cos^2(\theta)\right)=\\ = \left(\min(\mathcal{E}_{tot})+\cos^2(\theta)\right)
\left(\mathcal{E}_{tot}+\cos^2(\theta)\right);\nonumber
\end{eqnarray}
where $\min(\mathcal{E}_{tot})=\frac{(n^2-n_0^2)\cos^2\theta}{n^2-n_0^2+2n_0^2\cos^2\theta}$.

These two equalities can be used for an indirect measurement of the relative delays, $\mathcal{E}_{TE}$ and $\mathcal{E}_{TM}$, of the TE- and TM-polarized wave-packets by
experimental data obtained in a direct measurement of false relative group delay $\mathcal{E}_{tot} $. In addition, comparing the measured values of this quantity with its
values calculated on the basis of the expressions (\ref{80a}) and (\ref{44}), one can experimentally verify the timekeeping procedure itself.

\begin{figure}[h]
\begin{center}
\includegraphics[width=7.0cm]{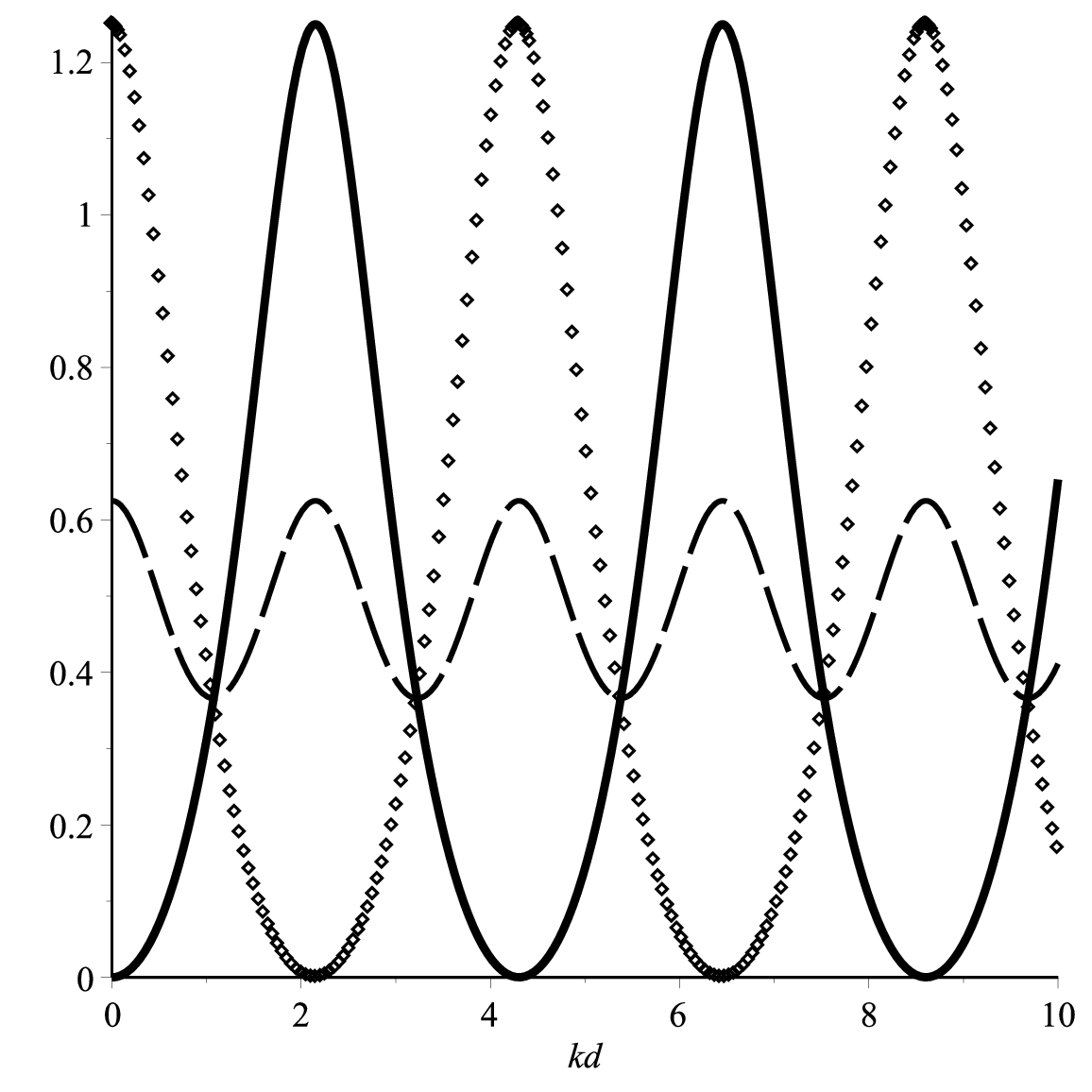}
\end{center}
\caption{The $kd$-dependence of $\mathcal{E}_{TE}$ (solid line), $\mathcal{E}_{TM}$ (points) and $\mathcal{E}_{tot}$ (dashed line) for $n_0=1$, $n=1.5$ and $\theta=20^o$.}
\label{fig.1}
\end{figure}

Fig. ~\ref{fig.1} shows the results of calculations $\mathcal{E}_{TE}$, $\mathcal{E}_{TM}$ and $\mathcal{E}_{tot}$ for $n_0=1$, $n=1.5$ and $\theta=20^o$.
The character of the dependence of these quantities on $kd$ is the same for all values of the parameter $\theta$ from the interval $[0,\pi/2)$. At the points where the functions
$\mathcal{E}_{TE}$ and $\mathcal{E}_{TM}$ take the minimum values, $T=1$. In this case, the corresponding group times $\tau_{TE}^{gr}$ and $\tau_{TM}^{gr}=\Delta\tau_{TM}^{gr}
\tau_{free}$ are $\tau_{free}$. That is, for $n>n_0$, the time $\tau_{free}$ of free passage through the layer never exceeds the group times $\tau_{TE}^{gr}$ and
$\tau_{TM}^{gr}$.

As for the case of $n<n_0$, both the times $\tau_{TE}^{gr}$ and $\tau_{TM}^{gr}$ can be abnormally short. However, it would be wrong to regard this fact as a defect of the
presented timekeeping procedure. The fact is that the case of $n<n_0$ does not correspond to the physical formulation of the problem -- for example, glass cannot fill all
infinite space outside the layer. The condition $n>n_0$ adopted in the section \ref{a1} is important. Moreover, even in the case of $n>n_0$, it would be correct to always
consider vacuum as a background medium. If we want to study the passage of light through the air gap between two semi-infinite spaces filled with glass, then we must consider
the problem of light scattering on a three-layer structure consisting of two layers of glass and an air gap between them (there should be a vacuum to the right and to the left
of this structure). Of course, in such a (quasi-one-dimensional) structure it is impossible to observe the regime of a frustrated total internal reflection. To study this mode,
it is necessary to consider a three-dimensional model with two glass prisms.

\section{On the known definitions of the group (phase) time} \label{Ch}

Note that the group times $\tau_{TE}^{gr}$ and $\tau_{TM}^{gr}$, introduced in this approach for TE- and TM-polarized wave packets, consisting of waves with a given angle of
incidence, differ from the known definitions of these quantities for the structure under study.

For example, the ``group-delay time"\/ $\tau_\gamma$ introduced in \cite{Ste} for scattering on the given structure of a two-dimensional wave packet (the waves forming such a
packet have not only different values of $k$, but also different values of $\theta$) is presented by Exp. (18), and the corresponding lateral shift $\Delta y$ is determined by
Eq. (16):
\begin{eqnarray*}
\tau_\gamma=\frac{1}{c}\left(J_k-\frac{\tan\theta}{k}J_\theta\right),\ooo \Delta y=-\frac{J_\theta}{k n_0\cos\theta}.
\end{eqnarray*}
That is, unlike ours, this version of the stationary phase approximation relates the lateral displacement with the derivative $J_\theta$, rather than with $J_k$.

Another ``group-delay time"\/, $\tau_S=J_k/c$, for this structure is presented in the review articles \cite{Shw} (see Exps. (1.1), (1.6) and (2.5)) and \cite{Win} (see Exp.
(34)). And it is important to emphasize that the corresponding lateral shift is introduced in \cite{Shw} (see (2.7), (2.9) and (2.10)) on the basis of the concept of the energy
flow velocity (related with the stationary scattering problem), rather than on the basis of the stationary phase approximation (related with the nonstationary scattering
problem).

As for $\tau_\gamma$, in this case $\tau_\gamma=\tau_{free}$. But $\tau_\gamma$ is not the time of passage of the wave packet maximum through the layer, since this quantity was
obtained without taking into account the fact that the peak of the falling packet $\mathcal{U}_{tot}$ does not turn into the peak of the transmitted packet. In addition, the
two-dimensional version of the \cite{Ste} method of the stationary phase has another serious defect. The fact is that it is based on the assumption that a small variation of the
angle of incidence $\Delta\theta$ for a wave packet with a small width $\Delta k$ in the space $k$ leads to a small change in its shape. However, it is not. Its width $\Delta
l\sim 1/\Delta k$ in ordinary space is large. And this means that the turn of the wave vector $\vec{k}$ on a small angle $\Delta\theta$ corresponds, on the scale of the packet
itself, to a far not small shift $\Delta l\cdot \Delta\theta\sim\Delta\theta/\Delta k$. Whence it follows that the double limit $\Delta\theta\to 0$ and $\Delta k\to 0$, the
existence of which is assumed by this version, does not really exist (it is not by chance that not only $\tau_S$, but also $\tau_\gamma$ can be less than $\tau_{free}$ and even
take negative values for $n_0 <n$). An internally consistent two-dimensional version of the stationary phase approximation (which implies the limit $\Delta k\to 0$) can be
developed only for the strict equality $\Delta\theta=0$. That is why namely this case was considered in this work.

\section{Conclusion}

The paper presents an alternative model of scattering of the plane monochromatic TE-wave on a uniform dielectric layer, in which the TE-wave incident on a layer is uniquely
represented as a superposition of two incident waves, one of which describes that component which is linked, on the plane of the mirror symmetry of the layer, with the
transmitted wave; while the other is associated with the reflected one. According to this model, the reflected component of the original TE-wave does not cross this plane. A
similar model has been developed for the TM-wave.

On the basis of these stationary models, the analysis of the scattering of TE- and TM-polarized, narrow  in the $k$-space, wave packets built of waves with the same angle of
incidence is carried out. For each polarization, by the example of the transmitted component, we have defined the (asymptotic) true and false group-delay times (the
corresponding values for the reflected and transmitted components coincide). Unlike the true group-delay time, the false group-delay time does not depend on the polarization of
the wave packet and allows direct measurement. Its name reflects the fact that this quantity describes the difference between two points in time that are not related to the
evolution {\ it of the same} wave packet. But just so, based on the concept of group velocity, the characteristic times are introduced for scattering on layered structures
within the framework of the standard approach. From the point of view of our approach, all such characteristics are 'false group times'.

So: (\i) a direct measurement of the true group-delay time for the layer is impossible in principle; (\i\i) the well-known phase time, which can be measured directly, is a false
group time; (\i\i\i) by experimental data obtained with help of direct measurements of the phase time it is possible to calculate (to indirectly measure), with making use the
equalities (\ref{88}), the true group-delay times for both polarizations.


\begin{thebibliography}{861}

\bibitem{Gha}
A. Ghatak and S. Banerjee, Appl. Optics. {\bf 28}, 1960 (1989)
\bibitem{Ste}
A. M. Steinberg and R. Y. Chiao, Phys. Rev. {\bf 49}, 3283 (1994)
\bibitem{Lee}
B. Lee and W. Lee, J. Opt. Soc. Am. B, {\bf 14}, 777 (1997)
\bibitem{Win}
Y. G. Winful, Phys. Rep. {\bf 436} 1 (2006)
\bibitem{Shw}
A. B. Shvartsburg, Physics-Uspekhi {\bf 50} (1) (2007)
\bibitem{Chu}
N. L. Chuprikov, Vestn. Samar. Gos. Tekhn. Univ., Ser. Fiz.-Mat. Nauki [J. Samara State Tech. Univ., Ser. Phys. Math. Sci.], 2013, Is. 2(31), P. 215--222
\bibitem{Born}
M. Born and E. Wolf, {\it Principles of optics}, Pergamon Press (1964)
\bibitem{But}
M. B\"{u}ttiker, R. Landauer, Phys. Rev. Lett. {\bf 49}, 1739 (1982)
\bibitem{But1}
M. Biittiker and R. Landauer, Phys. Scr. {\bf 32}, 429 (1985)

\end{thebibliography}
\end{document}